\documentstyle[twoside,fleqn,espcrc2,epsfig]{article}

\begin{document}
\begin{titlepage}
\setcounter{page}{0}
\topmargin 2mm
\begin{flushright}
{INPP-UVA-98-06}\\
\end{flushright}
\begin{center}
{\bf A Systematic Study of Power Corrections from World Deep Inelastic
Scattering Measurements.} 
\vskip 12pt
{Simonetta Liuti\footnote[1]{On leave from INFN, Sezione di Roma
Tre, via della Vasca Navale 84, I-00146 Rome, Italy}}\\
\vskip 6pt
{\em Institute of Nuclear and Particle Physics, University
of Virginia, \\ McCormick Road, Charlottesville, Virginia 22901, USA.}\\ 
\vskip 30pt
\end{center}
{\narrower\narrower
\centerline{\bf Abstract}
\medskip\noindent
By performing an analysis in moment space using high statistics DIS world 
data, we extract the values of both the QCD parameter 
$\Lambda^{(4)}_{\overline{MS}}$ 
up to NLO and of the power corrections to the proton structure function, 
$F_2$. At variance with previous analyses, the use of moments allows us 
to extend the kinematical range to 
larger values of $x$, where we find that power corrections are quantitatively
more important.
Our results are consistent with the
$n$ dependence predicted by IR renormalon calculations. We discuss 
preliminary
results on nuclear targets with the intent of illustrating a possible 
strategy to disentangle power corrections ascribed to IR renormalons 
from the ones generated  dynamically {\it e.g.}
from rescattering in the final state. The latter appear to be modified 
in nuclear targets.}

\smallskip
\vskip 60pt
\begin{center}
{Talk given at\\\smallskip {\em QCD98}\\ Montpellier, July 2nd-8th, 1998}\\
\smallskip
{\em to be published in the proceedings}\\
\end{center}
\bigskip
\vfill
\end{titlepage}

\title{A Systematic Study of Power Corrections from World Deep Inelastic
Scattering Measurements.}

\author{S. Liuti\address{Institute of Nuclear and Particle Physics, University
of Virginia, \\ McCormick Road, Charlottesville, Virginia 22901, USA.}%
        \thanks{On leave from: INFN-Sezione Roma Tre, Dipartimento
di Fisica E. Amaldi, Via Vasca Navale, 84. 00146 Roma, Italy.} }
       

\begin{abstract}
By performing an analysis in moment space using high statistics DIS world 
data, we extract the values of both the QCD parameter 
$\Lambda^{(4)}_{\overline{MS}}$ 
up to NLO and of the power corrections to the proton structure function, 
$F_2$. At variance with previous analyses, the use of moments allows us 
to extend the kinematical range to 
larger values of $x$, where we find that power corrections are quantitatively
more important.
Our results are consistent with the
$n$ dependence predicted by IR renormalon calculations. We discuss 
preliminary
results on nuclear targets with the intent of illustrating a possible 
strategy to disentangle power corrections ascribed to IR renormalons 
from the ones generated  dynamically {\it e.g.}
from rescattering in the final state. The latter appear to be modified 
in nuclear targets.    
\end{abstract}

\maketitle 

\section{QCD FITS TO MOMENTS OF $F_2$}

The $Q^2$ dependence of the
structure functions in Deep Inelastic Scattering (DIS) 
as well as in other related processes 
({\it e.g.} Drell-Yan) can be described accurately within perturbative QCD
(pQCD) up to NLO provided one 
confines calculations to the kinematical region of $Q^2 
\geq 10 \, {\rm GeV^2}$ and Bjorken $x$: $ 0.1 \leq x \leq 0.7$ (corresponding
to high values of the invariant mass squared: $W^2 \gg 4 \, {\rm GeV^2}$). 
At the border of this kinematical region 
the agreement between pQCD calculations and 
experiment is no longer accurate nor unambigous because 
of the increasing importance of non-perturbative (np) physics. 
At large $x$ the agreement with the data is known to improve
by including power corrections \cite{NMC,VM}. 
Aside from parton model based phenomenological interpretations, 
the presence of power corrections still constitutes a rather elusive
problem for theory.
Canonical methods such as the hard scattering factorization (HSF) or OPE 
(for a review see \cite{QiuSte} and references therein),  
are hampered by the large and hard-to-classify number of operators.
Comparison with data does not allow yet to distinguish among
different higher twist contributions (see {\it e.g.} \cite{Choi}). 
More recently the idea has been developed that  
a one to one correspondence can be defined between the dynamically
generated higher twist terms and the ambiguities 
in the resummation of the asymptotic pQCD series, or renormalons 
(\cite{Braun} and references therein). 
As for np methods, there 
seems to be indications from lattice studies 
that power corrections do surface 
though not seemingly directly related to renormalons \cite{Parri}.

A preliminary step in order to test recent developments is to
explore whether, 
given the accuracy and kinematical coverage of present data, 
power corrections can be separated unambigously and their 
actual size determined, from NLO and higher 
order terms, and from a number of other non-QCD contributions. 
With the aim of achieving a clear-cut answer, we started our analysis
from the most well known structure function, $F_2$.
\begin{table*}[hbt]
\setlength{\tabcolsep}{1.5pc}
%
\newlength{\digitwidth} \settowidth{\digitwidth}{\rm 0}
\catcode`?=\active \def?{\kern\digitwidth}
\caption{Comparison of world measurements of 
$\Lambda_{QCD} \equiv \Lambda^{(4)}_{\overline{MS}}$ and the  
coefficient $\tau^2$}
\label{results}
\begin{tabular}{|c|c|c|c|} \hline
Collaboration & Range in $Q^2$ {\rm $GeV^2$} &  $\Lambda^{(4)}_{\overline{MS}}$ MeV  &  $\tau^2$ {\rm $GeV^2$}  \\ \hline  
Before 1980 \cite{Penni}       &  $1.9 - 85$ &  $300 \leftrightarrow 1000$   &   $-0.1 \leftrightarrow 0.25$    \\ \hline
NMC \cite{NMC}       &  $0.5 - 260$ &  $ 250 $   &   $0.18 \pm 0.12$    \\ \hline
NMC+BCDMS+SLAC \cite{VM}  &  $0.5 - 260$  &  $263 \pm 42 \pm 55  $  & $0.09 \pm 0.04$  \\ \hline
%
%
CCFR \cite{CCFR} & $5 - 199.5$ &  $371 \pm 31 $ &  $0.24 \pm 0.1$  \\ \hline
BEBC-WA59 \cite{WA59}  &  $Q^2 < 64$ &  $110 ^{+50}_{-45}$ &  $-0.16 \pm 0.05$   \\ \hline
This paper $\gamma^4=0$    &  $5 - 260 $ & $ 241 \pm 36  $ &  $0.21 \pm 0.09$   \\ \hline
This paper $\gamma^4 \neq 0$   &  $2 - 260$ & $ 250 $ &  $0.25 \pm 0.20$   \\ \hline
\end{tabular}
\end{table*}

We constructed moments,
\begin{equation}
M_n(Q^2) = \int_0^1 dx F_2(x,Q^2) x^{n-2},
\label{mom}
\end{equation}
from the NMC, BCDMS and SLAC data sets. Differently from previous analyses,
we set no lower limit on the final state invariant mass, thus extending the
kinematical domain in $x$ up to $x \approx 1$ and 
$Q^2 \geq 2  \, {\rm GeV^2}$.   
Moments are more directly connected to OPE,
they can be calculated in principle using non-perturbative methods and, on
the practical side, they are the only way to include data at large $x$ 
without getting into the complications of the resonance structure.

We performed a fit of the following expression  
\begin{equation}
M_n(Q^2) = M_n^{pQCD} +
\sum_{k=1} M_n^{(2k+2)}(Q^2)\frac{1}{Q^{2k}},
\label{twist}
\end{equation}
where the $Q^2$ dependence of $M_n^{pQCD}= M_n(\mu^2) C_n(Q^2,\mu^2)$,
$\mu^2$ being a given scale, 
was considered up to NLO. Furthermore, since we are interested in the large $x$
behavior, we restricted the analysis to 
the Non Singlet (NS) contribution 
therefore avoiding ambiguities associated with the gluon distribution.
The subtraction of the Singlet contribution introduces some error. However,
based on current parametrizations we estimated it to 
be $7 - 12 \%$ of the total moment for $n=3$ and 
$Q^2= 2-50 \, {\rm GeV^2}$ and to be less then $1 \%$ for $n=8$. 
Further sources of systematic error arise from the extrapolation to regions
where no data are available which constitute $11.3 \%$ and $23.5 \%$
of the whole kinematic domain contributing to $M_n$ at 
$n=3$ and $n=8$, respectively, and from Target Mass Corrections (TMC)
which display an inverse power-like behavior and increase with $n$ \cite{Mir}. 
In order to be able to compare 
with previous extractions we will show here results
obtained in the ``factorized''  approximation, {\it i.e.}     
\begin{equation}
M_n(Q^2) = M_n^{pQCD} \left(1 + 
a_n^{(1)} \frac{\tau^2}{Q^2} + a_n^{(2)} \frac{\gamma^4}{Q^4}\right),
\label{fac}
\end{equation}
corresponding to $M_n^{(4)} \equiv a_n^{(1)} \tau^2$ etc.,  
where the $n$ dependence of the power corrections
is included in the functions $a_n^{(i)}$, $i=1,2$;  
$\tau^2$ and $\gamma^4$ are the parameters to be determined by the fit,
besides the pQCD ones; 
no $Q^2$ dependence is assumed for $M_n^{(4)}$.  
Power corrections of order higher than $O(1/Q^4)$ are found to be
negligible for $Q^2 \geq 2 \, {\rm GeV^2}$.
In order to compare with similar recent extractions \cite{VM,Choi},
we set in Eqs.(\ref{twist}) and (\ref{fac}): 
$\gamma^4 = 0$, and $\tau^2= M_n^{(4)}(Q^2)/n M_n^{pQCD}(Q^2)$.  
Our results can be also anti-Mellin transformed to reconstruct 
$F_2(x,Q^2)$, by using the properties of the Mellin transforms. 
Since the $n$ dependence found in different models is rather 
simple, no polynomial technique was found to be necessary at this level.

\section{DISCUSSION OF RESULTS}

Our results for the parameters $\Lambda^{(4)}_{\overline{MS}}$ and $\tau^2$ 
are given in Table \ref{results} and compared with
the ones obtained in previous extractions. Since recent analyses 
used directly $F_2(x,Q^2)$, giving $x$-dependent
power correction coefficients, $C_{HT}(x)$, we calculated the values of
$\tau^2$ shown in the Table by performing Mellin transforms 
of $C_{HT}(x)$, times
the correspondong pQCD structure functions parametrizations.
The form $a_n^{(1)}=n$, transforming into
$C_{HT}(x) \approx 1/(1-x)$ at large $x$ was used  
consistently.
  
The first 
line of Table \ref{results} was included 
in order to show the highly enhanced accuracy 
provided by the NMC, BCDMS and SLAC set of data
where it is now possible to perform 
a more careful extraction of power corrections.
An anti-correlation clearly emerges from our fits 
between the parameters of pQCD,
{\it e.g.} $\Lambda$, and the coefficients of the power corrections. 
We studied this correlation; details of our study are presented 
in \cite{Liuti}. Here it suffices to say that it originates from a region in 
$Q^2$ ($5 \leq Q^2 \leq 20 \, {\rm GeV^2}$) 
where neither the LO, simply logarithmic pQCD behavior, nor the
pure power correction terms are dominant, whereas a mixture of 
higher order and $1/Q^2$ terms are simultaneously present.

We dealt with this problem by extracting $\Lambda$ using only   
data at $Q^2 \geq 30 \, {\rm GeV^2}$ and by including   
lower values of $Q^2$ stepwise 
in the fit, at the fixed value of $\Lambda$. 
The value of $\Lambda$ we found is consistent with $250$ MeV 
{\it i.e.} the value previously found in \cite{NMC,VM}.        
The results presented in Table \ref{results} are therefore clear-cut 
in predicting that, for similar values of $\Lambda$, 
we find a contribution from power corrections
which is about {\em twice} as large
as in the previous QCD analysis. 
We seem however 
to find a contradiction with the analysis using neutrino data in 
that these show both a larger value of $\Lambda$ and a slightly enhanced 
value of $\tau^2$.  
By including the $O(1/Q^4)$ term and by using our approach of lowering
the $Q^2$ threshold from above, we found out that $\gamma^4 =0$ 
within error bars until one reaches the value of $Q_{min}^2 = 5 \, 
{\rm GeV^2}$.
Finally, $\tau^2$ and $\gamma^4$ are also anti-correlated.       
By including in the fit both parameters, $\tau^2$ and $\gamma^4$ with
the lowest allowed threshold, $Q_{min}^2 = 2 \, {\rm GeV^2}$, we are able to 
fit the data with negative values of $\gamma^4$ and  
$\tau^2 \approx 0.25 \, {\rm GeV^2}$,
{\it i.e.} consistent with the value obtained at a higher threshold. 
We would like to point out however that due to the anti-correlation between
the two parameters, by performing a general fit to the data regardless
of $Q^2$ ``thresholds'', large positive values of $\gamma^4$ and small values
of $\tau^2$ can be also found \cite{gangof4}; such results are to be 
interpreted as a manifestation of the anti-correlation between parameters.  
\begin{figure}[htb]
\vbox{
\hskip.6truecm\epsfig{figure=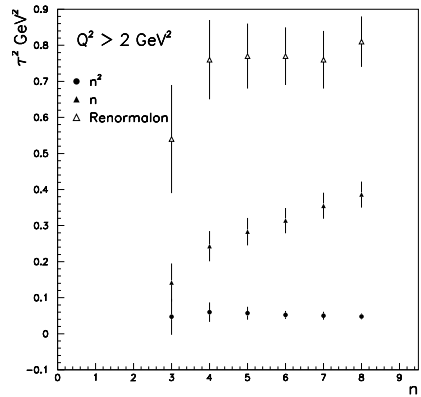,width=6.truecm}}
\vskip-.6truecm
\caption{The value of $\tau^2$ vs. $n$ 
extracted according to different models
for the $n$-dependent coefficient $a_n(1)$ as explained in the text.}
\label{fig1}
\end{figure}
%

With a firm understanding of the different $Q^2$ dependent contributions 
at intermediate and low $Q^2$, one can subsequently investigate the
renormalon hypothesis. We present here results of our study of the
$n$-dependence of the coefficient of the $1/Q^2$ term, $a_n^{(1)} \tau^2$,
in Eq.(\ref{fac}). In Figure 1 we repeated our fits 
using the forms: $a_n^{(1)} = n$, 
$a_n^{(1)} = n^2$, corresponding to the dominance of two-quark and four-quark
higher twist diagrams, respectively, in {\it e.g.} HSF models 
and  we compared them to fits using the $n$-dependence predicted by
renormalon calculations (for details see references [9] and [10] 
in \cite{Braun}).     
Since in all predictions the $n$-dependent term is factored out
from an unknown constant term, one can discriminate ``models''  based on
the ``flatness'' of the $n$ behavior extracted from the data.   
From Figure 1 one can see that the renormalon model gives clearly a better
interpretation than the widely used assumption, $a_n^{(1)}=n$.       
 
\section{CONCLUSIONS AND FUTURE DEVELOPMENTS}

Our initial study of power corrections in DIS is meant to be an exploratory 
one centered around two points: we first addressed the question of 
whether there is at all a direct and clear evidence of power corrections
in the data, and if
their actual magnitude can be extracted; we then moved on to a 
phenomeonological study of their nature, namely whether 
they can be associated to long distance features of pQCD 
(renormalon models) or if other dynamical mechanisms pertinent to 
DIS only, such as final state rescatterings should be taken into account.    
Our main conclusions on the first part are that indeed 
power corrections can be separated out from present data by using an 
approach in which different portions of the $Q^2$ domain (proceeding from
high to low $Q^2$ values) are analyzed separetely. The magnitude of the 
$O(1/Q^2)$ coefficient is $\tau^2 \approx 0.2 \, {\rm GeV^2}$, which will
need to be systematically compared with power corrections 
extracted from other reactions in further investigations, to test the
presence of a universal scale. An important outcome of our study is also
that, within the present accuracy and kinematical coverage of the data,
meaningful results should be presented including the correlation between 
the different parameters of the fit. Our extractions seem to be consistent 
with renormalon calculations. 

More detailed knowledge on the nature 
of power corrections can be sought in principle 
by using nuclear targets.             
By assuming the validity of IA, {\it i.e.} by disregarding
nucleon rescattering effects, one obtains a factorization of the
$A$ and $Q^2$ dependences 
of the moments \cite{CDL}  
\begin{equation}
M_n^A(Q^2)  =  {\cal M} _n^A \,  M_n^N(Q^2), 
\label{momA}
\end{equation}
Deviations from such a factorization would correspond to
to a breakdown of IA and they would therefore signal the presence 
of rescattering effects. As shown in \cite{CDL} present data 
on nuclear targets are not accurate enough to observe any 
feature in the ratio of moments and they support the IA hypothesis.
More detailed studies, together with higher accuracy nuclear DIS data
might be able to reveal an underlying  structure \cite{Liuti}.
At present, by adopting the 
factorized form, Eq.(\ref{momA}), 
we can extract the value of $\tau^2$
for a nucleon. Results are presented in Figure 2 and compared with 
the proton values. From the Figure we can see that the proton and nucleon 
results are consitent with one another, {\it i.e.} no difference 
is seen in the neutron contribution.     
 
\begin{figure}[htb]
\vbox{
\hskip.6truecm\epsfig{figure=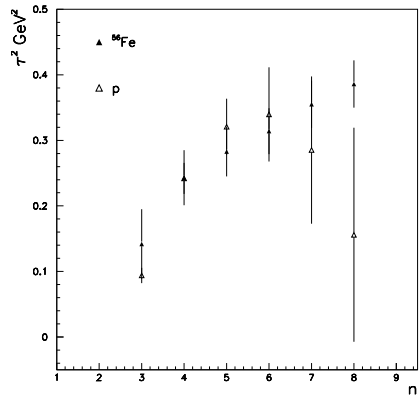,width=6.truecm}}
\vskip-.8truecm
\caption{$\tau^2$ vs. $n$ 
extracted from $^{56}Fe$ and proton data.}
\label{fig2}
\end{figure}

I thank the Institute of Nuclear and Particle Physics at the 
University of Virginia for support and the Argonne National Laboratory 
Nuclear Theory Group where part of this contribution was written. 
I also thank Lech Mankiewicz for discussions.

\end{document}